\documentclass[onecolumn]{article}
\usepackage{amsmath}

\newtheorem{theorem}{Theorem}
\newtheorem{acknowledgement}[theorem]{Acknowledgement}

\input{tcilatex}

\begin{document}

\title{Decoherence and purity of a driven solid-state qubit in Ohmic bath}
\author{Zhi-Qiang Li, Xian-Ting Liang\thanks{%
Corresponding author. E-mail: xtliang@ustc.edu} \\
Department of Physics and Institute of Modern Physics,\\
Ningbo University, Ningbo 315211, China}
\maketitle

\begin{abstract}
In this paper we study the decoherence and purity of a driven solid-state
qubit in the Ohmic bath by using the method based on the\ master equation.
At first, instead of solving the master equation we investigate the
coefficients of the equation which describe the shift in frequency,
diffusive, decoherence, and so on. It is shown that one of the coefficients
(we called it decoherence coefficient) is crucial to the decoherence of the
qubit in the model. Then we investigate the evolution of the purity of the
state in the model. From the analysis of the purity we see that the
decoherence time of the qubit decrease with the increase of the amplitude of
the driven fields and it is increase with the increase of the frequency of
the driven fields.


\end{abstract}

\section{Introduction}

Solid-state qubits are considered to be promising candidates for realizing
building blocks of quantum information processors. In past few years, many
kinds of these qubit models are introduced and many efforts not only
theoretical but also experimental have been performed on investigating the
decoherence, relaxation and manipulation for these qubit models. Most of the
investigations for these models are considered them as a spin-boson model
which is modeled with a two-level system coupled to a bath and the bath
always constructed with a set of hamonic oscillators. The qubits can be
manipulated with a driven field \cite{Nature (London) 398 (1999) 786,Science
296 (2002) 886,Science 296 (2002) 889,Phys. Rev. Lett. 89 (2002)
117901,Nature (London) 406 (2000) 43}, so a more realistic description
requires the inclusion of an external control field in the qubit-bath model.
The quantum tunneling or other quantum properties of the driven two-level
system have been investigated at dept in last years \cite{rep 304
(1998)229,Adv. Phys. 54 (2005) 525}. Recently, the driven spin-boson model
attracting a lot of attention because of its interest in connection with
quantum computing with solid state devices. The control of the coherence
dynamics of a two-level atom placed in a lossy cavity with an external
periodic driving field \cite{j.mod.opt.47.2905}, the dynamics of a XOR gate
operation with an external source \cite{pra65.012309}, and the consequence
of driving in terms of multi-phoyo transtions experimentally \cite{prl
93.037001} are investigated. In this paper, we shall analyze the influence
of the driven control field on the decoherence of the qubits by using the
master equation method \cite{W. H.
Zurek(PRD)24-1516(1981),WH(PRD)26-1862(1982),J.p.pazPRD45(1992)2843,J.P.Paz
W.H. PRL 82(1999)5181}. At first we shall derive out a master equation from
the Hamiltonian of the driven qubit in the Ohmic bath. Then we analyze the
decoherence and other decay coefficients based on the master equations.
Finally, a rate of purity decay for this model is obtained and discussed.

\section{Models and Master Equation}

In this paper we concentrate on the case of a persistent current qubit based
on Josephson junction \cite{science 285(1999)1036,science 290(2000)
773,science 299(2003) 1869,nature 431(2004)159}. It is driven with the
magnetic flux through the loop and damped predominantly by flux noise with
Gaussian statistics. This setup is accurately expressed by the driven
spin-boson (DSB) model \cite{rev.mod.phys.59 1, Prb 68 (2003) 012508, pre 61
(2000) r4687,rep 304 (1998)229}%
\begin{equation}
H=H_{S}+H_{B}+H_{I},  \label{1}
\end{equation}%
where%
\begin{equation}
\begin{array}{c}
H_{s}=\frac{1}{2}\varepsilon (t)\hat{\sigma}_{z}-\frac{1}{2}\Delta \hat{%
\sigma}_{x}, \\ 
H_{B}=\sum_{n}(\frac{P_{n}^{2}}{2m_{n}}+m_{n}\omega
_{n}^{2}x_{n}^{2}),H_{I}=\sigma _{z}\sum_{n}\lambda _{n}x_{n}.%
\end{array}
\label{2}
\end{equation}%
Here, $\sigma _{i}$ ($i=x,y,z$) are Pauli matrices. The quantum environment
is modeled with an infinite set of harmonic oscillators of mass $m_{n}$,
angular frequency $\omega _{n}$, momentum $p_{n}$ and position coordinate $%
x_{n}$ which are coupled independently to the spin $\sigma _{z}$ with
strength measured by the set \{$\lambda _{n}$\} and $\varepsilon
(t)=\varepsilon _{0}+s\cos (\Omega t)$ is the external, time-dependent
control field with the static basis $\varepsilon _{0}$. For $s=0$, the
system degenerates to be a common two-level one.

The evolution of the total density matrix for the system, in the interaction
picture, reads%
\begin{equation}
i\hbar \overset{\dot{\sim}}{\rho }_{T}=[\overset{\sim }{H_{I}},\overset{\sim 
}{\rho }_{T}],  \label{3}
\end{equation}%
where the interaction representation of the operators are given by 
\begin{equation*}
\overset{\sim }{\rho }_{T}(t)=\exp (\frac{iH_{0}t}{\hbar })\rho _{T}\exp (%
\frac{-iH_{0}t}{\hbar }),\ddot{o}
\end{equation*}%
\begin{equation}
\overset{\sim }{H_{I}}(t)=\exp (\frac{iH_{0}t}{\hbar })H_{I}\exp (\frac{%
-iH_{0}t}{\hbar }).  \label{4}
\end{equation}%
Here $\rho _{T}=\rho _{T}\left( 0\right) $ and $H_{I}=H_{I}\left( 0\right) $%
, where $H_{I}\left( 0\right) $ is the Hamiltonian of the system in the Schr%
\"{o}dinger picture. The perturbative expression of the Eq. (\ref{4}) in the
interaction picture is given to second order of $H_{I}$ by%
\begin{eqnarray}
\overset{\sim }{\rho _{T}}(t) &=&\overset{\sim }{\rho _{T}}(0)+\frac{1}{%
i\hbar }\int_{0}^{t}dt_{1}[\overset{\sim }{H_{I}}(t_{1}),\overset{\sim }{%
\rho _{T}}(0)]  \notag \\
&&+(\frac{1}{i\hbar })^{2}\int_{0}^{t}dt_{1}\int_{0}^{t_{1}}dt_{2}[\overset{%
\sim }{H_{I}}(t_{1}),[\overset{\sim }{H_{I}}(t_{2}),\overset{\sim }{\rho _{T}%
}(0)]].  \label{5}
\end{eqnarray}%
The reduced density operator for the system is defined by $\overset{\sim }{%
\rho }(t)=$Tr$_{B}(\overset{\sim }{\rho _{T}}(t))$, Where Tr$_{B}$ indicates
a trace over environment variables, then we assume that initially the system
and environment are uncorrelated that the total density matrix is a tensor
product of the form $\overset{\sim }{\rho _{T}}(t)=\overset{\sim }{\rho }%
(0)\otimes \overset{\sim }{\rho _{B}}(0)$. So we get 
\begin{eqnarray}
\overset{\sim }{\rho }(t) &=&\overset{\sim }{\rho }(0)+\frac{1}{i\hbar }%
\int_{0}^{t}dt_{1}\text{Tr}[\overset{\sim }{H_{I}}(t_{1}),\overset{\sim }{%
\rho }(0)\otimes \overset{\sim }{\rho _{B}}(0)]  \notag \\
&&+(\frac{1}{i\hbar })^{2}\int_{0}^{t}dt_{1}\int_{0}^{t_{1}}dt_{2}\text{Tr}[%
\overset{\sim }{H_{I}}(t_{1}),[\overset{\sim }{H_{I}}(t_{2}),\overset{\sim }{%
\rho }(0)\otimes \overset{\sim }{\rho _{B}}(0)]].  \label{6}
\end{eqnarray}%
Next, we make a rather trivial operation that enables us to finish the
derivation in a simple way, we can express the initial state $\overset{\sim }%
{\rho }(0)$ in terms of $\overset{\sim }{\rho }(t)$ using the same
perturbative expansion and rewrite the Eq. (\ref{5}) while the initial state 
$\overset{\sim }{\rho }(0)$ appears in the right-hand side only, then
inserting this expression into Eq. (\ref{6}) and making the derivation \cite%
{J. P.Paz W.H.Zurek book}, we can obtain%
\begin{eqnarray}
\overset{\dot{\sim}}{\rho }(t) &=&\frac{1}{i\hbar }\int_{0}^{t}dt_{1}\text{Tr%
}[\overset{\sim }{H_{I}}(t_{1}),\overset{\sim }{\rho }\otimes \overset{\sim }%
{\rho _{B}}]-\frac{1}{\hbar ^{2}}\int_{0}^{t}dt_{1}\text{Tr}[\overset{\sim }{%
H_{I}}(t),[\overset{\sim }{H_{I}}(t_{1}),\overset{\sim }{\rho }\otimes 
\overset{\sim }{\rho _{B}}]]  \notag \\
&&+\frac{1}{\hbar ^{2}}\int_{0}^{t}dt_{1}\text{Tr}[\overset{\sim }{H_{I}}(t),%
\text{Tr}[\overset{\sim }{H_{I}}(t_{1}),\overset{\sim }{\rho }\otimes 
\overset{\sim }{\rho _{B}}]\otimes \overset{\sim }{\rho _{B}}].  \label{7}
\end{eqnarray}%
The Eq. (\ref{7}) is in the interaction picture. By virtue of the evolution
operator $U_{s}=\exp [-\frac{i}{\hbar }\int_{0}^{t}dt_{1}H_{s}(t_{1})]$ we
rewrite the equation in the Schr\"{o}dinger picture, and considering the
coupling term in the DSB with $H_{I}=\sigma _{z}\sum_{n}\lambda _{n}x_{n}$,
then we get the master equation of DSB in the Schr\"{o}dinger picture is%
\begin{eqnarray}
\dot{\rho} &=&\frac{1}{i\hbar }[H_{eff},\rho ]+\frac{1}{i\hbar }%
\sum_{n}[\lambda _{n}\left\langle x_{n}\right\rangle \sigma _{z},\rho ] 
\notag \\
&&-\frac{1}{\hbar ^{2}}\sum_{n}\int_{0}^{t}dt_{1}(k_{n}^{(1)}[\sigma
_{z},[\sigma _{z}(t_{1}-t),\rho ]]+k_{n}^{(2)}[\sigma _{z},\{\sigma
_{z}(t_{1}-t),\rho \}]),  \label{8}
\end{eqnarray}%
with 
\begin{eqnarray}
k_{n}^{(1)} &=&\frac{1}{2}\lambda _{n}^{2}\left\langle
\{x_{n}(t),x_{n}(t_{1})\}\right\rangle -\lambda _{n}^{2}\left\langle
x_{n}\right\rangle \left\langle x_{n}\right\rangle ,  \notag \\
k_{n}^{(2)} &=&\frac{1}{2}\lambda _{n}^{2}\left\langle
[x_{n}(t),x_{n}(t_{1})]\right\rangle ,  \label{9} \\
H_{eff} &=&a_{1}\hat{\sigma}_{x}+b_{1}\hat{\sigma}_{y}+c_{1}\hat{\sigma}_{z},
\notag
\end{eqnarray}%
where,%
\begin{eqnarray}
a_{1} &=&\frac{\Delta }{2}\cos (\frac{\varepsilon _{0}t+\frac{s}{\Omega }%
\sin (\Omega t)}{\hbar })+\frac{1}{4\hbar }\left[ \Delta (\varepsilon _{0}t+%
\frac{s}{\Omega }\sin (\Omega t))\right.  \notag \\
&&\left. +\Delta t(\varepsilon _{0}+s\cos (\Omega t))\right] \times \cos (%
\frac{\Delta t}{\hbar })\sin (\frac{\varepsilon _{0}t+\frac{s}{\Omega }\sin
(\Omega t)}{\hbar }),  \notag \\
b_{1} &=&\frac{\Delta }{2}\sin (\frac{\varepsilon _{0}t+\frac{s}{\Omega }%
\sin (\Omega t)}{\hbar })-\frac{1}{4\hbar }\left[ \Delta (\varepsilon _{0}t+%
\frac{s}{\Omega }\sin (\Omega t))\right.  \notag \\
&&\left. +\Delta t(\varepsilon _{0}+s\cos (\Omega t))\right] \times \cos (%
\frac{\Delta t}{\hbar })\cos (\frac{\varepsilon _{0}t+\frac{s}{\Omega }\sin
(\Omega t)}{\hbar }),  \notag \\
c_{1} &=&-\frac{\Delta }{2}(\varepsilon _{0}+s\cos (\Omega t))+\frac{1}{%
4\hbar }\left[ \Delta (\varepsilon _{0}t+\frac{s}{\Omega }\sin (\Omega
t))\right.  \notag \\
&&\left. +\Delta t(\varepsilon _{0}+s\cos (\Omega t))\right] \sin (\frac{%
\Delta t}{\hbar }).  \label{10}
\end{eqnarray}

Here, the notation [$\cdot $,$\cdot $], \{$\cdot $,$\cdot $\} denote the
commutators and anticommutators, and $\left\langle ..\right\rangle =$Tr$%
(..\rho )$ are reserved for quantum expectation values. So far, we only make
two important assumptions: first, we used a perturbative expansion up to
second order, which is accurate enough as the interaction small enough
comparing to the Hamiltonians of system and bath; second, we assumed that
the initial state is not entangled. Next we consider another important
approximation, which is usually considered in the master equation, the
Markov approximation. Assuming the kernel $k_{n}^{(i)}$ are strongly peaked
at the point $t=t_{1}$, with slowly varying functions, then we can transform
the temporal integrals over the variable $\tau =t-t_{1}$. And assumed that
the initial state of the environment to be thermal equilibrium at
temperature $T=1/k_{B}\beta $, thus the first order term in the master
equation of DSB disappears because $\left\langle x_{n}\right\rangle =0.$
Therefore, the master equation becomes%
\begin{equation}
\dot{\rho}=\frac{1}{i\hbar }[H_{eff},\rho ]-\frac{1}{\hbar }%
\int_{0}^{t}dt_{1}(\nu (t_{1})[\sigma _{z},[\sigma _{z}(-t_{1}),\rho
]]-i\eta (t_{1})[\sigma _{z},\{\sigma _{z}(-t_{1}),\rho \}]),  \label{11}
\end{equation}%
with the two kernels are the dissipation and noise kernels respectively and
are defined as%
\begin{equation}
\nu (t)=\frac{1}{2\hbar }\sum_{n}\lambda _{n}^{2}\left\langle \left\{
x_{n}(t),x_{n}(0)\right\} \right\rangle =\int_{0}^{\infty }d\omega J(\omega
)\cos (\omega t)\coth (\frac{\beta \hbar \omega }{2}),  \label{12}
\end{equation}%
\begin{equation}
\eta (t)=\frac{1}{2\hbar }\sum_{n}\lambda _{n}^{2}\left\langle
[x_{n}(t),x_{n}(0)]\right\rangle =\int_{0}^{\infty }d\omega J(\omega )\sin
(\omega t),  \label{13}
\end{equation}%
where $J(\omega )$is the bath spectral density function defined by%
\begin{equation}
J(\omega )=\frac{\pi }{2}\sum \frac{\lambda _{n}^{2}}{m_{n}\omega _{n}^{2}}%
\delta (\omega -\omega _{n}).  \label{14}
\end{equation}%
We consider that the Hamiltonian of the spin system $H_{s}=\frac{1}{2}%
\varepsilon (t)\hat{\sigma}_{z}-\frac{1}{2}\Delta \hat{\sigma}_{x}$ , the
evolution operator has the form 
\begin{eqnarray}
U_{s} &=&\exp [-\frac{i}{\hbar }\int_{0}^{t}dt_{1}H_{s}(t_{1})]  \notag \\
&=&\exp \{-\frac{i}{2\hbar }[\varepsilon _{0}t+\frac{s}{\Omega }\sin (\Omega
t)]\hat{\sigma}_{z}\}\exp \{\frac{i}{2\hbar }\Delta t\hat{\sigma}_{x}\} 
\notag \\
&&\times \exp \{-\frac{i\Delta }{(2\hbar )^{2}}[\varepsilon _{0}t^{2}+t\frac{%
s}{\Omega }\sin (\Omega t)]\hat{\sigma}_{y}+o(t^{3})\},  \label{15}
\end{eqnarray}%
where we using the formula $e^{A+B}=e^{A}e^{B}\exp \sum_{j}\sum_{i}\frac{1}{%
j!}\frac{1}{i!}\frac{1}{j+i+1}[[B,A^{(i)}],B^{(j)}]$. If we set the time $t\ 
$very small (namely, $t$ is smaller than the characteristic time of the
qubit), we can ignore the term of $o(t^{3})$. So we can solve the Heisenberg
equations for the system and determine the operator $\sigma _{z}(t)$ to be 
\begin{eqnarray}
\sigma _{z}(t) &=&-\sigma _{x}\cos (\frac{\Delta t}{\hbar })\sin (\frac{%
\Delta t}{2\hbar ^{2}}[\varepsilon _{0}t+\frac{s}{\Omega }\sin (\Omega
t)])-\sigma _{y}\sin (\frac{\Delta t}{\hbar })  \notag \\
&&+\sigma _{z}\cos (\frac{\Delta t}{\hbar })\cos (\frac{\Delta t}{2\hbar ^{2}%
}[\varepsilon _{0}t+\frac{s}{\Omega }\sin (\Omega t)]).  \label{16}
\end{eqnarray}%
So the results of this paper are all the short-time results. Substituting
this equation into Eq. (\ref{11}), we obtain the final expression for the
master equation of DSB. It is 
\begin{eqnarray}
\dot{\rho} &=&\frac{1}{i\hbar }[H_{eff}+\tilde{\Omega}(t),\rho ]-D(t)[\sigma
_{z},[\sigma _{z},\rho ]]-G(t)[\sigma _{z},[\sigma _{x},\rho ]]  \notag \\
&&-f(t)[\sigma _{z},[\sigma _{y},\rho ]]\text{ }+ir_{1}(t)[\sigma
_{z},\{\sigma _{x},\rho \}]+ir_{2}(t)[\sigma _{z},\{\sigma _{y},\rho \}]
\label{17}
\end{eqnarray}%
where%
\begin{equation*}
D(t)=\int_{0}^{t}dt_{1}\nu (t_{1})(\cos (\frac{\Delta t_{1}}{\hbar })\cos (%
\frac{\Delta t_{1}}{2\hbar ^{2}}[\varepsilon _{0}t_{1}+\frac{s}{\Omega }\sin
(\Omega t_{1})])),
\end{equation*}%
\begin{equation*}
f(t)=\int_{0}^{t}dt_{1}\nu (t_{1})(-\sin (\frac{\Delta t}{\hbar })),
\end{equation*}%
\begin{equation*}
G(t)=\int_{0}^{t}dt_{1}\nu (t_{1})(-\cos (\frac{\Delta t}{\hbar })\sin (%
\frac{\Delta t}{2\hbar ^{2}}[\varepsilon _{0}t+\frac{s}{\Omega }\sin (\Omega
t)])),
\end{equation*}%
\begin{equation}
\tilde{\Omega}(t)=\int_{0}^{t}dt_{1}\eta (t_{1})(\cos (\frac{\Delta t_{1}}{%
\hbar })\cos (\frac{\Delta t_{1}}{2\hbar ^{2}}[\varepsilon _{0}t_{1}+\frac{s%
}{\Omega }\sin (\Omega t_{1})])),  \label{18}
\end{equation}%
\begin{equation*}
r_{1}(t)=\int_{0}^{t}dt_{1}\eta (t_{1})(-\cos (\frac{\Delta t}{\hbar })\sin (%
\frac{\Delta t}{2\hbar ^{2}}[\varepsilon _{0}t+\frac{s}{\Omega }\sin (\Omega
t)])),
\end{equation*}%
\begin{equation*}
r_{2}(t)=\int_{0}^{t}dt_{1}\eta (t_{1})(-\sin (\frac{\Delta t}{\hbar })).
\end{equation*}

\section{Decoherence and purity decay}

Having obtained a master equation describing the evolution of a controllable
solid state qubit system coupled to an Ohmic bath, we now proceed to examine
the consequences of that evolution. In the following we shall analyze the
problem with two different methods. At first, we use the coefficients in the
master equation to detect the instantaneous effects of the environment. Then
we can investigate the time evolution of the purity. For the first method,
all the effects including controlled external field and uncontrolled
environment are considered in the above coefficients. These coefficients are
to renormalize the frequency as well as to introduce the decay of the
system. From these equations, it is possible to have a qualitative idea of
the effects for the environment producing on the system. First, we can
observe that the effective Hamiltonian $H_{eff}$ is evolved with
time-dependence. The system Hamiltonian $H_{S}$ and the coupling Hamiltonian 
$H_{I}$ do not commute with each other and the evolution of $H_{eff}$ is
periodic because the driven controlled field is periodic. The term $\tilde{%
\Omega}(t)$ is the shift in frequency which produces the renomalized
frequency. This term does not affect the unitarity of the evolution. The
terms $D(t)$, $f(t)$, $G(t)$, $r_{1}(t)$ and $r_{2}(t)$ are diffusive terms
and bring about non-unitary effects. The terms of $r_{1}(t)$ and $r_{2}(t)$
are the dissipation coefficients related to the dissipation kernel $\nu (t)$
defined already, which play the role of a time-dependent relaxation rate,
are independent of temperature. And $D(t)$, $f(t)$ and $G(t)$ are the
diffusion coefficients, which produce the decoherence effects, they are not
only time-dependent but depend on temperature. Of cause, the explicit time
dependence of the coefficients can only be computed once we specify the
spectral density of the environment. To illustrate their qualitative
behavior, we focus on the case that the environment is the Ohmic bath in the
following, whose spectral density of the Ohmic bath can be expressed as 
\begin{equation}
J(\omega )=2\pi \alpha \omega \exp (-\frac{\omega }{\Lambda }),  \label{19}
\end{equation}%
where $\Lambda $ is the physical high-frequency cutoff, which represents the
highest frequency presented in the environment and the parameter $\alpha $
is dimensionless parameter reflecting the strength of dissipation. We set $%
\alpha =0.01$ as in \cite{PRE 61 r4687}.\ We can plot the diffusive terms $%
D(t)$, $f(t)$ and $G(t)$ as Fig.1.%
\begin{equation*}
Fig.1
\end{equation*}%
It is shown that all these plots appear to be a periodically diverging
function, which stems not only from\ the periodic control driven field but
also from the bath, and the peaks of $D(t)$ is much higher than that of $%
f(t) $ and $G(t)$. It means that the term proportional to $D(t)$ in Eq. (\ref%
{18}) plays the main role to decoherence. The form of the decoherence
coefficient $D(t)$ \cite{J. P.Paz W.H.Zurek book} is not necessarily
intuitive, so we evaluate it numerically as Fig.2. 
\begin{equation*}
Fig.2
\end{equation*}%
One of the parameters ($s,$ $T,$ $\Omega $) is varied in Figs.2 (a, b, c).
All of the plots demonstrate the same basic behavior: $D(t)$ decreases with
periodic oscillations.

It is shown that the influence of the heat bath to the decoherence of the
qubit can be described by these coefficients of the master equation. It can
also be described by a computationally convenient way, namely the linear
entropy $\zeta =tr(\rho -\rho ^{2})$ which is another measure of the purity
of a quantum state. For the pure state it is approximative equals to $\zeta
=1-tr(\rho ^{2})$ so we shall study the evolution of the purity of the
system as measured by $\xi =tr(\rho ^{2})$ for simplicity \cite{J. P.Paz
W.H.Zurek book, PRL 70 1187 (1993),pra 68(2003)032104}. It is equal to one
for a pure state and decreases when the state of the system gets mixed
because the destruction of quantum coherence is generated by evolution. By
virtue of the master equation, we can easily obtain an evolution for the
purity $\xi $ and the equation we obtained is:%
\begin{eqnarray}
\dot{\xi} &=&-4D(t)Tr(\rho ^{2}\sigma _{z}^{2}-\rho \sigma _{z}\rho \sigma
_{z})-4G(t)Tr(\rho \sigma _{x}\rho \sigma _{z})-f(t)Tr(\rho \sigma _{y}\rho
\sigma _{z})  \notag \\
&&-4r_{1}(t)Tr(\rho ^{2}\sigma _{y})+4r_{1}(t)Tr(\rho ^{2}\sigma _{x}).
\label{20}
\end{eqnarray}%
Setting the initial state is pure then $\rho ^{2}=\rho $. Similar to Ref. %
\cite{Forschr.phys.54.804(2006)} we can obtain 
\begin{equation}
\dot{\xi}(t)=-\frac{4}{3}D(t).  \label{21}
\end{equation}%
It is interesting that the Eq. (\ref{21}) only correspond to the decoherence
coefficient $D(t)$. It is shown that the purity decay rate is proportional
to the decoherence coefficient $D(t)$, and do not include any other
coefficients. It confirms that $D(t)$ plays the main role in decoherence
which is analyzed in the above section in detail. From Eq.(\ref{21}) we can
also see that the larger the rate of the purity is, the smaller the
decoherence becomes. In Fig.3, we plot the evolutions of the state purity
with time $t$.

\begin{equation*}
fig.3
\end{equation*}%
It is show that the purity of qubit decreases fast with the increasing of
the amplitude and decreasing of the frequency of the driven control field.
This is agree with the result in Ref.\cite{Forschr.phys.54.804(2006)} that
the low frequency driven field is destructive to the coherence of the qubit.

\section{Conclusions}

In this paper we have investigated the effects of the environment for a
solid state qubit coupled with a driven control field. At first, instead of
solving the master equation we investigated the coefficients in the equation
which express the shift in frequency, diffusive, decoherence and so on. Here
we suppose the environment is modeled with a Ohmic bath and it is coupled
with the qubit linearly. It shows that the term proportional to the
coefficient $D(t)$ in master equation plays the main role to the
decoherence. Then, we concretely investigated the decay of purity. It is
shown that the decay rate of purity behaves fast as the amplitude increasing
and the frequency decreasing of the driven field. It is suggested that if
the driven field is necessary for some qubit system a higher frequency may
help to decrease the decoherence.

\begin{acknowledgement}
This project was sponsored by National Natural Science Foundation of China
(Grant No. 10675066) and K. C. Wong Magna Fundation in Ningbo University.
\end{acknowledgement}

\section{Figures Captions}

Fig. 1: Coefficients in master equation versus time $t$. (a) $D(t)$; (b) $%
f(t)$; (c) $h(t)$. Here, the parameters are $\varepsilon _{0}=\Delta $, $%
s=\Delta $, $\Omega =\Delta $, $\Lambda =36\Delta $, $\Delta $ $=5$, $T=30$
mK, here and in the following figures the times are expressed in unit of $%
5/\Delta $.

Fig.2:\ Dependence of $D(t)$ on coefficients of the driven field and
temperature $T$. The parameters $s,$ $\Omega ,$ $\beta ,$ and $\Delta $ are
same as Fig. 1. (a) $D(t)$ versus $s$; (b) $D(t)$ versus $\Omega $; (c) $%
D(t) $ versus $\beta =1/kT$.

Fig.3: Dependence of purity of the system on coefficients of the driven
field. The parameters $s,$ $\Omega ,$ $\beta ,$ and $\Delta $ are same as
Fig. 1. (a) $\xi $ versus $s$; (b) $\xi $ versus $\Omega $.


\begin{thebibliography}{99}
\bibitem{Nature (London) 398 (1999) 786} Y. Nakamura, Yu. A. Pashkin, and J.
S. Tsai, Nature (London) 398 (1999) 786.

\bibitem{Science 296 (2002) 886} D. Vion, A. Aassime, A. Cottet, P. Joyez,
H. Pothier, C. Urbina, D. Esteve, and M. H. Devoret, Science 296 (2002) 886.

\bibitem{Science 296 (2002) 889} Y. Yu, S. Han, X. Chu, S.-I. Chu, and Z.
Wang, Science 296 (2002) 889.

\bibitem{Phys. Rev. Lett. 89 (2002) 117901} J. M. Martinis, S. Nam, J.
Aumentado, and C. Urbina, Phys. Rev. Lett. 89 (2002) 117901.

\bibitem{Nature (London) 406 (2000) 43} J. R. Friedman, V. Patel, W. Chen,
S. K. Tolpygo, J. E. Lukens, Nature (London) 406 (2000) 43.

\bibitem{rep 304 (1998)229} M. Grifoni, P. H\"{a}nggi, Phys. Rep. 304 (1998)
229.

\bibitem{Adv. Phys. 54 (2005) 525} I. Goychuk, and P. Hanggi, Adv. Phys. 54
(2005) 525.

\bibitem{science 290(2000) 773} C. H. van der Wal, A. C. J. ter Haar, F. K.
Wilhelm, R. N. Schouten, C. J. P. M. Harmans, T. P. Orlando, S. Lloyd, J. E.
Mooij,\ Science 290(2000) 773.

\bibitem{science 285(1999)1036} J. E. Mooij, T. P. Orlando, L. Levitov, L.
Tian, C. H. van der Wal, S. Lloyd, Science 285 (1999) 1036.

\bibitem{science 299(2003) 1869} I. Chiorescu, Y. Nakamura, C. J. P. M.
Harmans, J. E. Mooij, Science 299 (2003) 1869.

\bibitem{nature 431(2004)159} I. Chiorescu, P. Bertet, K. Semba, Y.
Nakamura, C. J. P. M. Harmans, J. E. Mooij, Nature 431 (2004) 159.

\bibitem{W. H. Zurek(PRD)24-1516(1981)} W. H. Zurek, Phys. Rev. D 24 (1981)
1516.

\bibitem{WH(PRD)26-1862(1982)} W. H. Zurek, Phys. Rev. D 26 (1982) 1862.

\bibitem{rev.mod.phys.59 1} A. J. Leggett, S. Chakravarty, A. T. Dorsey, M.
P. A. Fisher, A. Garg, W. Zwerger, Rev.Mod. Phys. 59, 1 (1987).

\bibitem{Prb 68 (2003) 012508} M. C. Goorden, F. K. Wilhelm, Phys. Rev. B 68
(2003) 012508.

\bibitem{pre 61 (2000) r4687} L. Hartmann, I. Goychuk, M. Grifoni, P. H\"{a}%
nggi, Phys. Rev. E 61 (2000) R4687.

\bibitem{PRL 70 1187 (1993)} W. H. Zurek, S. Habib, J. P. Paz, Phys. Rev.
Lett 70 (1993) 1187.

\bibitem{J.P.Paz  W.H. PRL 82(1999)5181} J. P. Paz, W. H. Zurek, Phys. Rev.
Lett 82 (1999) 5181.

\bibitem{J.p.pazPRD45(1992)2843} B. L. Hu, J. P. Paz, Y. H. Zhang, Phys.
Rev. D 45 (1992) 2843.\ 

\bibitem{J. P.Paz W.H.Zurek book} J. P. Paz, W. H. Zurek, \emph{%
Environment-induced decoherence and the transition from quantum to classical}%
, (Springer Berlin/Heidelberg) (2002).

\bibitem{pra 68(2003)032104} R. B. Kohout, W. H. Zurek, Phys. Rev. A 68
(2003) 032104.

\bibitem{Forschr.phys.54.804(2006)} S. Kohler, P. H\"{a}nggi, Forschr. Phys.
54 (2006) 804.

\bibitem{PRE 61 r4687} L. Hartmann, I. Goychuk, M. Grifoni, P. H\"{a}nggi.
Phys. Rev. E 60 (2000) R4687.

\bibitem{j.mod.opt.47.2905} M. Thorwart, L. Hartmann, I. Goychuk, P. H\"{a}%
nggi.J. Mod. Opt. 47 (2000) 2905.

\bibitem{pra65.012309} M. Thorwart, P. H\"{a}nggi. Phys. Rev. A 65 (2002)
012309.

\bibitem{prl 93.037001} S. Saito, M. Thorwart, H. Tanaka, M. Ueda, H.
Nakano, K. Semba, and H. Takayanagi. Phys. Rev. Lett. 93 (2004) 037001.
\end{thebibliography}
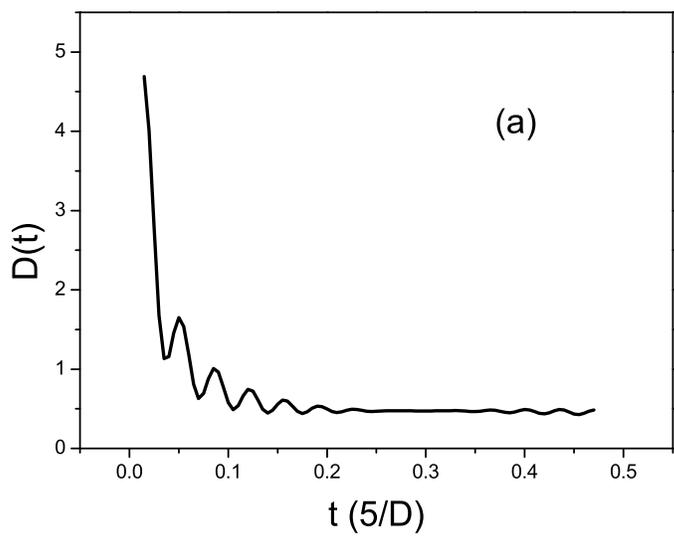
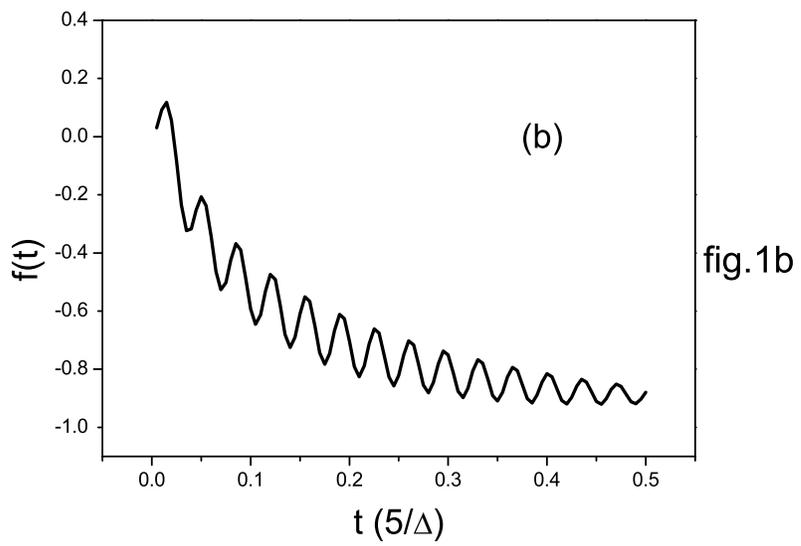
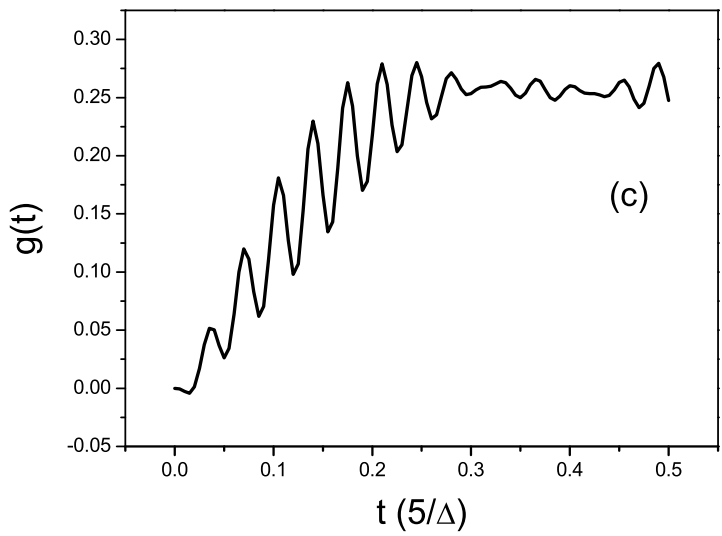
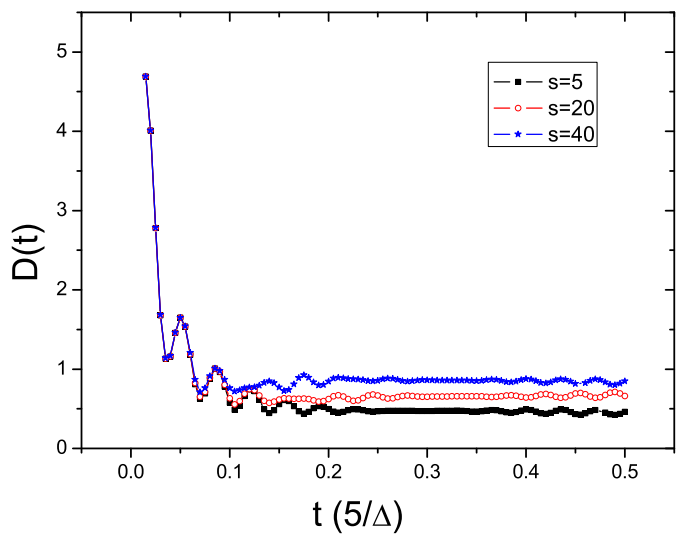
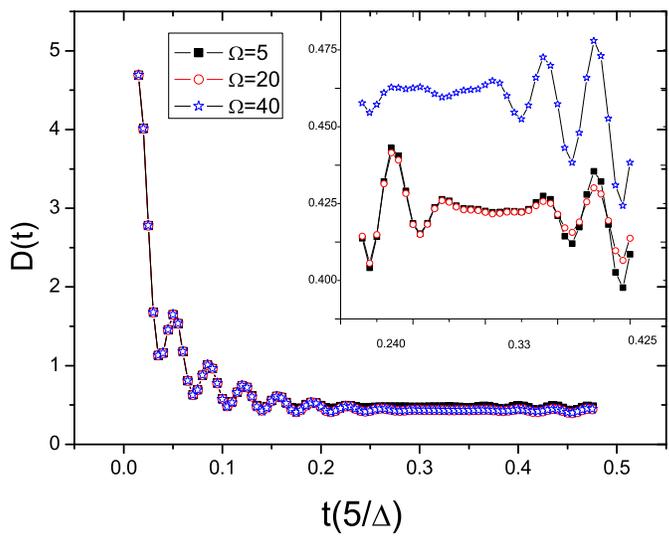
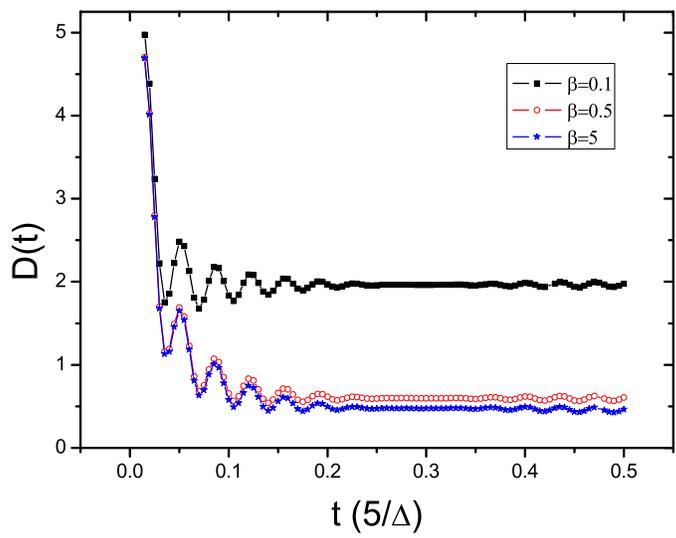
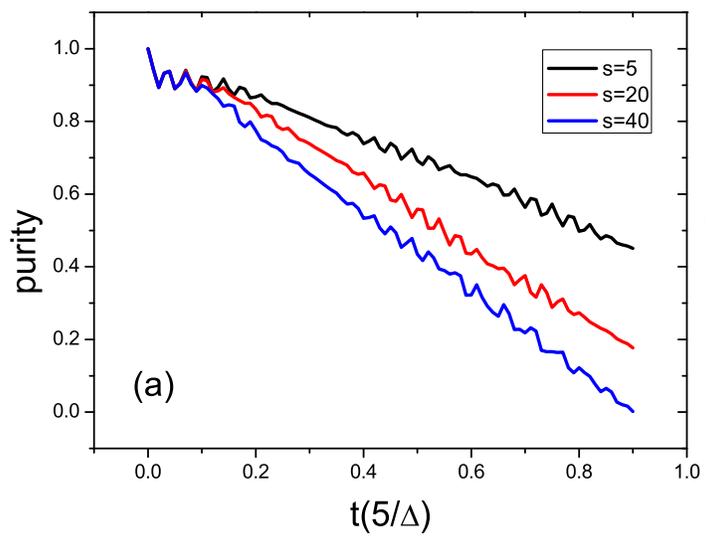
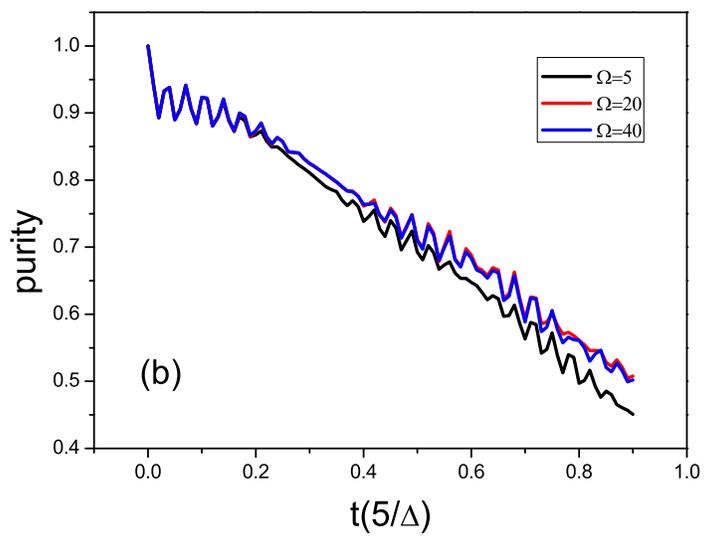
\end{document}